\documentclass[aps,showpacs,prb,twocolumn,amsmath,amssymb]{revtex4}
\usepackage{amssymb}
\usepackage{graphicx}% Include figure files
\usepackage{dcolumn}% Align table columns on decimal point
\usepackage{bm}% bold math

\begin{document}

\bibliographystyle{apsrev}

\newcommand{\bs}{{\bm \sigma}}
\newcommand{\bt}{{\bm \tau}}
\newcommand{\BS}{Bi$_2$Se$_3$}
\newcommand{\BX}{Bi$_2$X$_3$}

\newcommand{\la}{\langle}
\newcommand{\ra}{\rangle}
\newcommand{\da}{\dagger}
\newcommand{\up}{\uparrow}
\newcommand{\down}{\downarrow}

\newcommand{\lc}{\lowercase}

\newcommand{\no}{\nonumber}
\newcommand{\be}{\begin{equation}} 
\newcommand{\ee}{\end{equation}}
\newcommand{\bea}{\begin{eqnarray}} 
\newcommand{\eea}{\end{eqnarray}}

\newcommand{\vQ}{{\bf Q}}
\newcommand{\vk}{{\bf k}}
\newcommand{\vq}{{\bf q}}
\newcommand{\vM}{{\bf M}}
\newcommand{\vm}{{\bf m}}
\newcommand{\vB}{{\bf B}}
\newcommand{\vh}{{\bf h}}
\newcommand{\vvr}{{\bf r}}
\newcommand{\vu}{{\bf u}}
\newcommand{\hz}{\hat{{\bf z}}}
\newcommand{\hk}{\hat{k}}
\newcommand{\hq}{\hat{q}}
\newcommand{\hB}{\hat{B}}
\newcommand{\ham}{\hat{m}}
\newcommand{\hmu}{\hat{\mu}}
\newcommand{\tm}{\tilde{m}}
\newcommand{\tB}{\tilde{B}}
\newcommand{\tk}{\tilde{k}}
\newcommand{\hl}{\hat{\lambda}}
\newcommand{\hP}{\hat{\Phi}}

\newcommand{\Q}{($0,0,\frac{2\pi}{c}$)}

%%%%%%% Title %%%%%%%%%%%%%%%%%%%%%%%%%%%%%%%%%%%%%%%%%%%%%%%%%%%%%%%%%%%%%

\title{Magnetic torque oscillations from warped helical surface states\\ in topological insulators}

%%%%%%% Author(s) %%%%%%%%%%%%%%%%%%%%%%%%%%%%%%%%%%%%%%%%%%%%%%%%%%%%%%%%%
\author{Peter Thalmeier}
\affiliation{Max Planck Institute for Chemical Physics of Solids,
01187 Dresden, Germany}
\date{\today}
%%%%%%% Abstract %%%%%%%%%%%%%%%%%%%%%%%%%%%%%%%%%%%%%%%%%%%%%%%%%%%%%%%%%%%%%%
\begin{abstract}

A magnetic torque method is proposed that probes the warping and mass gap of Dirac cone surface states  in topological insulators like  \BX~ (X=Se,Te). A rotating field parallel to the surface induces a paramagnetic moment in the helical surface states for nonzero warping. It is non-collinear with the  applied field and therefore produces torque oscillations as function of the field angle which are a direct signature of the  surface states. The torque dependence on field strength and angle, the chemical potential and the Dirac cone parameters like warping strength and mass gap is calculated. It is shown that the latter leads to a symmetry reduction in the fourfold torque oscillations.

\end{abstract}


\maketitle

%%%%%%%%%%%%%%%%% %%%%%%%%%%%%%%%%%%%%%%%%%%%%%%%%%%%%%%%%%%%%%%%%%%%%%%%%%

\section{Introduction}
\label{sect:introduction}

The strong topological insulators \cite{hasan:10} of the \BX~(X=Se,Te) family, including the various doped systems  are characterized by 
surface states comprising a single Dirac cone at the $\Gamma$ point of the Brillouin zone (BZ) \cite{hsieh:09}. These states are protected against the
effect of disorder, e.g., impurity doping as long as time reversal invariance is preserved \cite{fu:07,chen:10a,wray:10,pan:11}. In the simplest effective models the surface states have a helical spin polarization orthogonal to the wave vector on the circular Fermi surface (FS) which is cut out from the cone at the position of the  chemical potential. However, it has become clear that this picture is oversimplified. When the chemical potential is moved away from the Dirac point the cone becomes warped and the FS changes from circular to hexagonal snowflake shape. This effect has been directly observed in photoemission \cite{chen:09} and explained as the result of higher order terms in the effective surface Hamiltonian \cite{liu:10,zhang:09,zhou:09,fu:09}. Furthermore ferromagnetic order of Fe dopants may lead to the opening of a mass gap in the Dirac cone due to breaking of time reversal invariance \cite{chen:10a,wray:10}.

In experiments it is not always easy to discriminate the contribution of surface states from those of the bulk \cite{checkelsky:10,butch:10}. In this work we propose a straightforward and sensitive magnetic torque method to observe the physical effect of surface states and signatures  of their cone structure. The schematic setup is shown in the inset of Fig.~\ref{fig:Fig2}. A magnetic field \vB~ is applied and rotated parallel to the surface of a disk-like sample. The c-axis is oriented perpendicular to the plane. When the chemical potential is away from the bottom and top of bulk bands a paramagnetic moment due to polarization of helical spins of surface states may be induced. Note that in this  geometry there is no diamagnetic effect, i.e., no Landau quantization of Dirac cones. However the spin polarization leads to a moment \vM~ which is non-collinear to \vB, giving rise to a magnetic torque $\bt_s=V_{SL}\vM\times\vB=\tau_S\hz$ where $V_{SL}$ is the total volume of the surface sheets on both sides of the sample. The torque may be measured via piezoelectric elements  fixed to the disk (inset in Fig.~\ref{fig:Fig2}). In fact recently this method has been successfully applied to measure the tiny torque oscillations from a collective hidden order parameter in a low carrier heavy fermion system \cite{okazaki:11}. A theoretical analysis of this effect was given in Ref. ~\onlinecite{thalmeier:11b}. In the present case the torque oscillation is entirely due to the virtual particle-hole continuum excitations of the warped  Dirac cone.  We propose this method as a promising way to investigate the surface states. Furthermore the torque amplitude is sensitive to the chemical potential and the opening of a mass gap at the Dirac point caused by magnetic surface doping (e.g. by Fe \cite{chen:10a,wray:10}). When the moments are perpendicular to the surface there will also be an in-plane torque ($\perp \hz$) originating directly from the magnetic surface layer. However, it is  independent of the direction of \vB~ and also independent of the chemical potential. Because it is perpendicular to the torque from Dirac surface states it should be possible to discriminate it from the latter by an experimental setup as shown in Fig.~\ref{fig:Fig2}.

In Sec.~\ref{sect:torque} we derive the expression for the surface torque based on an effective low energy model for surface states. The discussion of numerical results is presented in Sec.~\ref{sect:numerical} and the conclusion is given in Sec.~\ref{sect:conclusion}. \\ 
%
%%%%%%%%%%%%%%%%%%%%%%%%%%%%%%%%%%%%%%%%%%%%%%%%%%%%%%%%%%%%%
\begin{figure}
%\vspace{0.2cm}
\includegraphics[width=42mm]{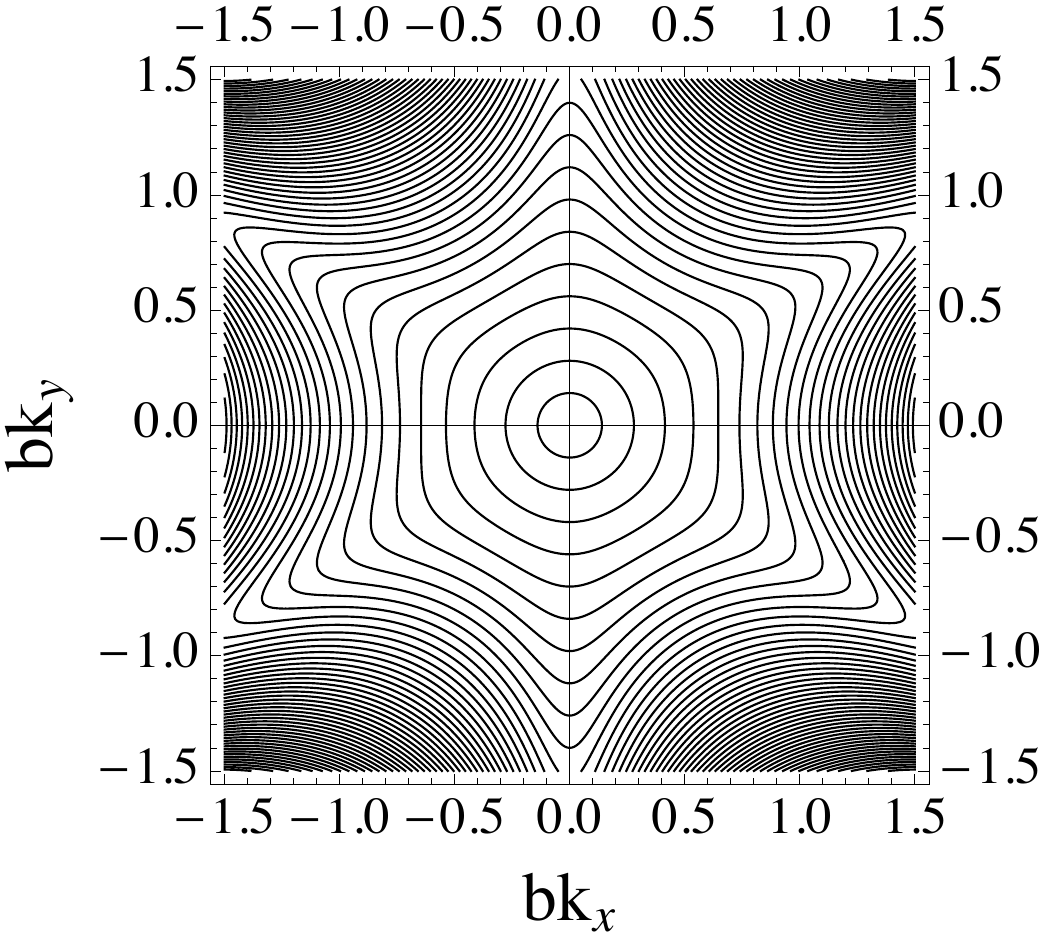}\hfill
\includegraphics[width=42mm]{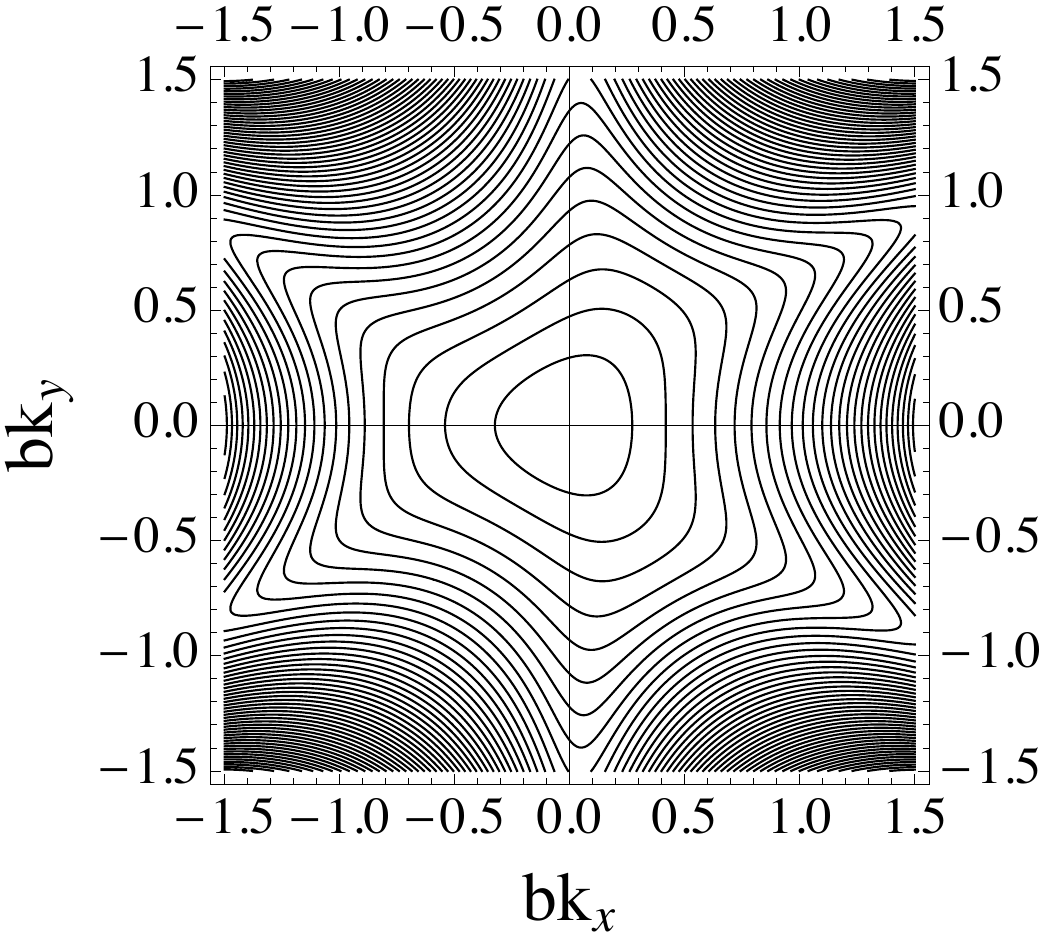}\\
\includegraphics[width=42mm]{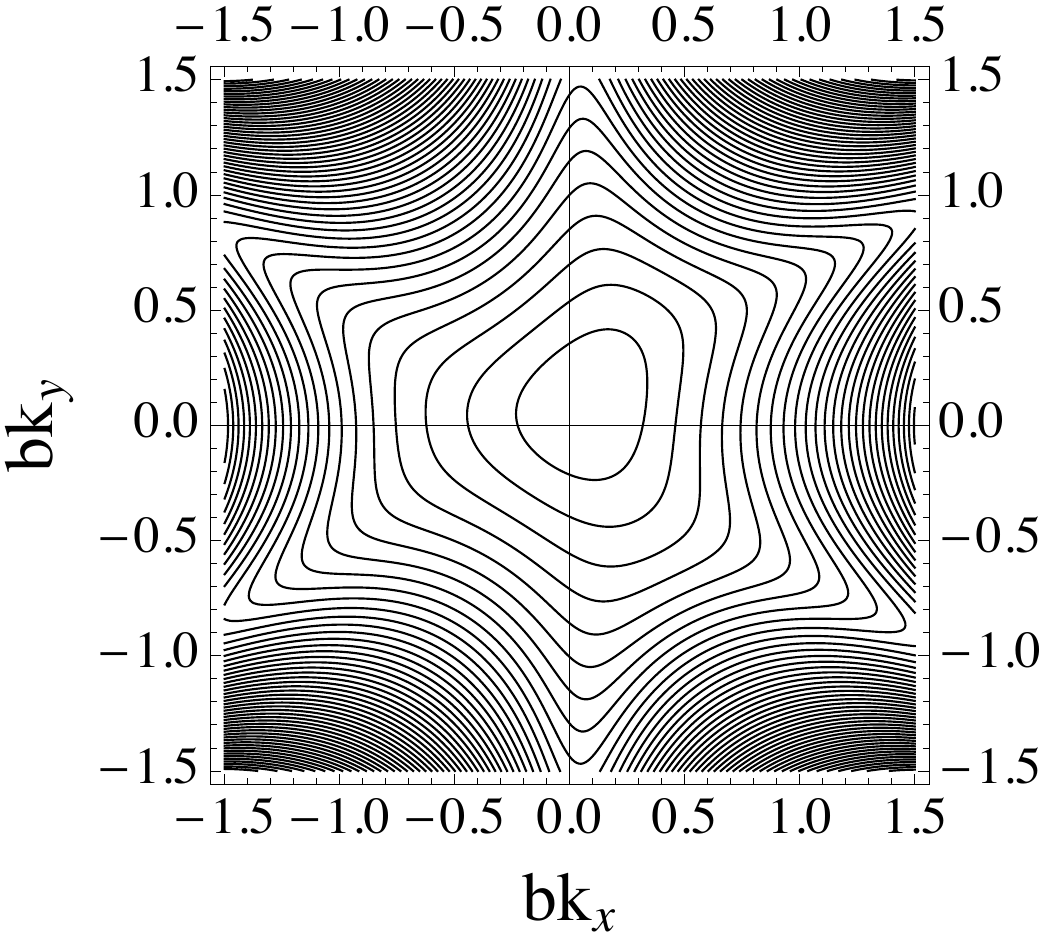}\hfill
\includegraphics[width=42mm]{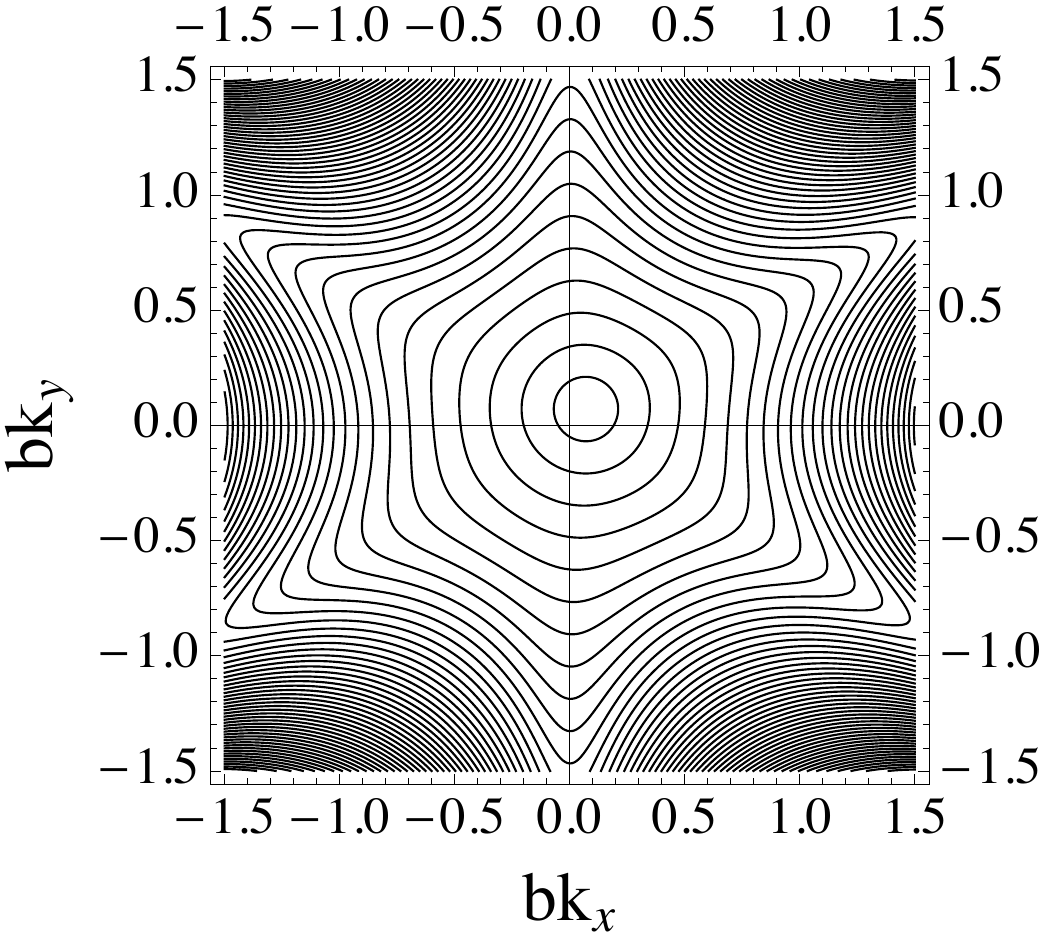}
\caption{Contour plots of warped and gapped Dirac cones in an in-plane magnetic field ($\phi=3\pi/4$). To be compatible with Eqs.~(\ref{eq:dispk},\ref{eq:warp}) we use the convention of $k_x,k_y$ corresponding to $\Gamma K, \Gamma M$ directions in the hexagonal surface BZ respectively.
Here intrinsic units $(\tk_x,\tk_y) = (bk_x, bk_y)$ (Table \ref{table:tableI}) are used with $\tilde{m}=m/E^*; \tilde{B}=\gamma_SB/E^*$  where the energy scale is $E^*=v_F/b=(v_F^3/\lambda)^\frac{1}{2}$. In clockwise direction (a-d): $(\tilde{m},\tilde{B})= (0,0); (0.3,0); (0, 0.1); (0.3, 0.1)$. The mass term (chosen large for clarity)  reduces the rotational symmetry while the field shifts the Dirac or extremal point of the dispersion.}
\label{fig:Fig1}
\end{figure}
%%%%%%%%%%%%%%%%%%%%%%%%%%%%%%%%%%%%%%%%%%%%%%%%%%%%%%%%%%%%%
%

 \section{Magnetic torque and paramagnetic susceptibility in the warped Dirac cone model}
 \label{sect:torque}
 
There are two contributions to the total out-of plane torque originating from the bulk ($\tau_B$) and topological surface states ($\tau_S$) respectively. The latter is obtained from the paramagnetic susceptibility tensor $\tensor{\chi}^S$ of surface states by $\vM=\tensor{\chi}^S\vB$. Then we can express the surface torque contribution as \cite{thalmeier:11b}
\bea
\tau_S(\phi)=\frac{1}{2}V_{SL}B^2[\Delta\chi^S(\phi)\sin 2\phi -2\chi_{xy}^S(\phi)\cos 2\phi]
\label{eq:torque1}
\eea
with $\Delta\chi^S=\chi_{xx}^S-\chi_{yy}^S$ and $V_{SL}=2A\delta_S$ (A= disk area, $\delta_S$=surface state localisation length which is of the order of a quintuple layer thickness in \BX). Here $\phi$ is the azimuthal field angle , i.e., $\vB=B(\cos\phi,\sin\phi)$.  The question arises whether $\tau_S$ can be separated from the bulk part. The latter is due to the occupied valence band states. If the chemical potential varies within the bulk band gap, then $\tau_B$ will be constant whereas $\tau_S$ will strongly depend on the chemical potential as discussed in Sec.~\ref{sect:numerical}. Therefore if the difference of $\tau=\tau_B+\tau_S$ at chemical potentials $\mu$ and $-\mu_c$ (inset of Fig.~\ref{fig:Fig3}) is considered, then the surface part may be extracted by $\tau_S(\mu)=\tau(\mu)-\tau(-\mu_c)$ since $\tau_S(-\mu_c$)=0. The relative magnitude of $\tau_S$ to $\tau_B$ cannot be calculated in the present effective low energy surface state model used below. A rough estimate may be obtained from the following consideration: If we take a thin-film like sample composed of n quintuple layers of \BX~ and assume that the surface states are localised within one unit one may estimate $\tau_S/\tau_B\sim 2/(n-2)$ for $\mu\simeq 0$. The surface state Dirac cones exist for $n>5$ (Ref.~\onlinecite{he:10}), then in the optimum case one has $\tau_S/\tau_B\simeq 2/3$. Even for $n\simeq 10^2$ $\tau_S$ would still be 2\% of $\tau_B$. Since torques can be measured with relative accuracy of $10^{-4}$ (Ref. \onlinecite{okazaki:11}) there seems to be no real obstacle to extract the surface torque contribution. Naturally one should try to observe this effect in thin films as function of the layer number similar to the photoemission experiments for samples with $n< 50$ (Ref.~\onlinecite{he:10}) and also to thin film transport measurements \cite{kim:11}.
%
%%%%%%%%%%%%%%%%%%%%%%%%%%%%%%%%%%%%%%%%%%%%%%%%%%%%%%%%%%%%%
\begin{figure}
\vspace{0.2cm}
\includegraphics[width=84mm]{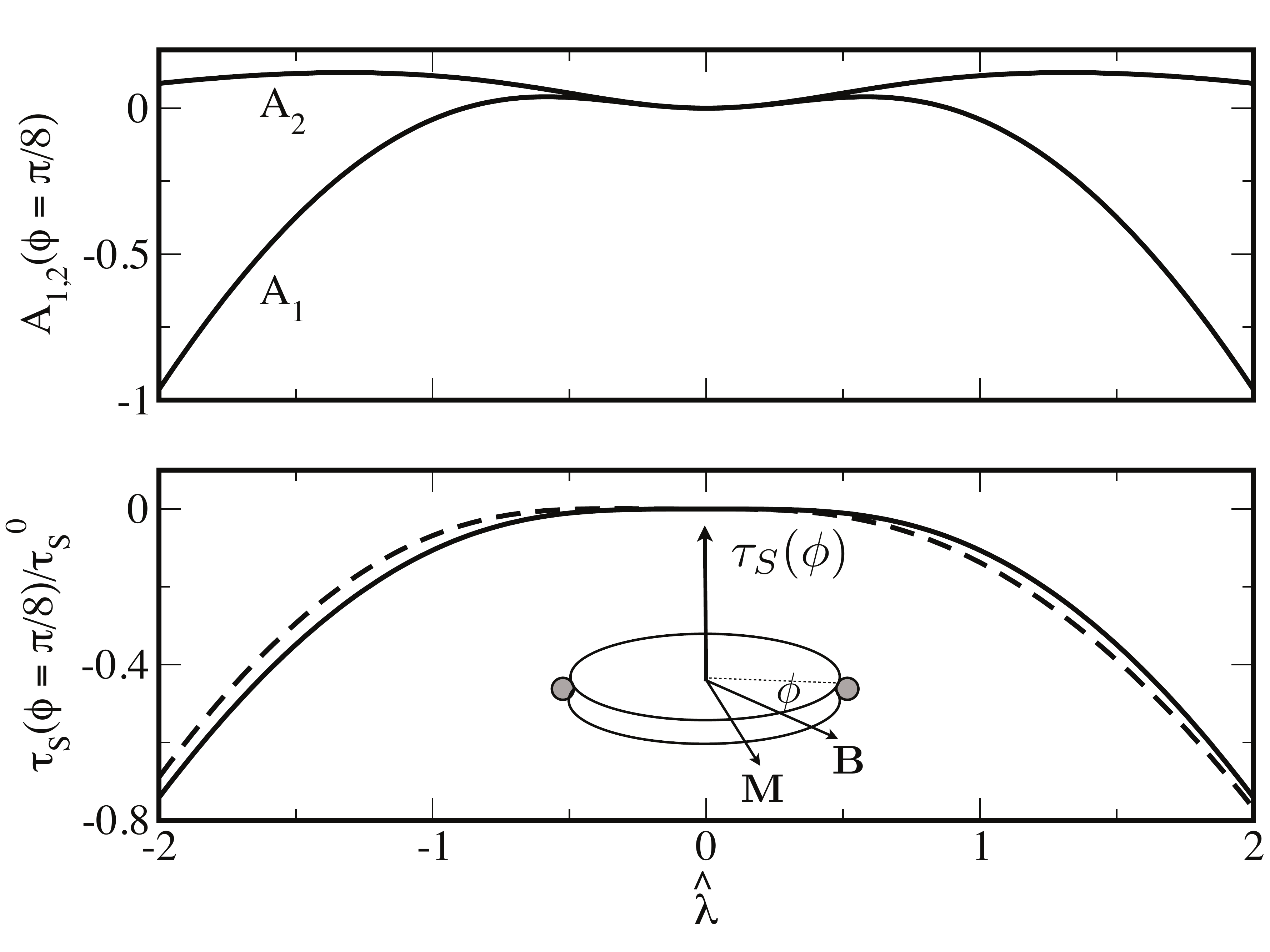}
\caption{Top: Variation of torque amplitudes $A_{1,2}$ for $\hat{\mu}=0.2, \hB =0.1$ at $\phi=\pi/8$  with (dimensionless) Dirac cone warping parameter $\hl=\lambda q_c^3/\mu_c$, here $\ham = 0$.  
Bottom: Total torque $\hat{\tau}_S$ for $\ham =0$ (full line) and  $\ham =0.1$ (dashed line). For finite $\ham$ the symmetry under sign change of $\hl$ is lost. Inset shows schematic geometry with disk-like sample fixed by piezoelectric elements (grey circles) to measure the torque $\bt_S(\phi)$ under rotating field \vB. }
\label{fig:Fig2}
\end{figure}
%%%%%%%%%%%%%%%%%%%%%%%%%%%%%%%%%%%%%%%%%%%%%%%%%%%%%%%%%%%%%
%

To identify the effect of surface states on the torque we now have to calculate the field angular dependence of the susceptibility tensor caused by the surface states. An effective low energy surface model may be obtained by starting from four atomic spin-orbit and crystalline electric field split orbitals of quintuple layers in \BX~  with appropriate surface boundary conditions \cite{liu:10}. The form of the model Hamiltonian  \cite{zhang:09}  is dictated by inversion symmetry I , time reversal T and the threefold rotation $C_3$. Explicitly one obtains:
\bea
H(\vk)&=&E_{0\vk} + v_F(k_x\sigma_y-k_y\sigma_x)
+\frac{1}{2}\lambda(k_+^3+k_-^3)\sigma_z \no\\
&&+m\sigma_z -\gamma_S(B_x\sigma_x+B_y\sigma_y)
\label{eq:ham}
\eea
where $k_\pm=k_x\pm ik_y$. The first parabolic term $E_{0\vk}=\vk^2/2m^*$ does not contribute to $\Delta\chi^S$ or $\chi_{xy}^S$ and is therefore irrelevant for the torque as is shown below. The second term describes the perfect Dirac cone and the third term is the third order warping part whose strength is given by the constant $\lambda$. Similar fifth order terms will be neglected. The fourth term is due to T reversal symmetry  breaking  caused e.g. by ferromagnetically ordered Fe surface doping. Assuming Ising type easy axis order (moment $\parallel\hz$) this leads to the opening of a mass gap at the Dirac point \cite{chen:10a,nomura:10,mong:10}.  This may be derived from a RKKY mechanism including all spin components \cite{liu:09}.
Finally the last term is the Zeeman term with $\gamma_S=\frac{1}{2}g_S\mu_B$ where $g_S$ is the effective surface state g-factor which is in the range \cite{analytis:10} $g_S\simeq 50$. We note that under reflection $\sigma_h$ with respect to the
horizontal disk center plane the Hamiltonian transforms as $\sigma_h[H(\vk)]=H(-\vk)$. Therefore surface states on both sides of the disk contribute additively to \vM. We assume here that the disk is sufficiently thick to prevent coupling between surface states on both sides. The evolution of surface states as function of thickness was considered in detail in Ref.~\onlinecite{hao:11}.
The warped Dirac cone dispersion is obtained from Eq.~(\ref{eq:ham}) as
\bea
\label{eq:dispk}
\epsilon_{\vk\tau}&=&E_{0\vk}+\tau(\epsilon_{D\vk}^2+\epsilon_{W\vk}^2)^\frac{1}{2}\no\\
\epsilon_{D\vk}^2&=&(v_Fk_x-\gamma_SB_y)^2+(v_Fk_y+\gamma_SB_x)^2\\
\epsilon_{W\vk}^2&=&(m+\lambda\Phi_\vk)^2\no
\eea
where we defined the warping function as 
\bea
\Phi_{\vk}=\frac{1}{2}(k_+^3+k_-^3)=k_x(k_x^2-3k_y^2)
\label{eq:warp}
\eea
Contour plots of this dispersion in the $(k_x,k_y)$ plane that show the effect of warping , finite mass and Zeeman term are presented in Fig.~\ref{fig:Fig1} where intrinsic units $\tm,\tB$ defined in the caption are used. The characteristics of the dispersion will be discussed below. It is obvious that $\epsilon_{D\vk}$ is the gapless {\it isotropic} Dirac cone dispersion shifted in the $(k_x,k_y)$- plane to a new origin by the shift vector 
\bea
\vk_s=\frac{\gamma_S}{v_F}(B_y,-B_x)=\frac{\gamma_SB}{v_F}(\sin\phi,-\cos\phi)
\eea
In the following index 's' refers to shifted quantities. Defining the momentum $\vq=\vk-\vk_s$ with respect to the shifted orgin the dispersion in Eq.~(\ref{eq:dispk}) may be rewritten in the form
\bea
\epsilon^s_{\vq\tau}&=&E^s_{0\vq}+\tau\epsilon^s_\vq\no\\
&=&E^s_{0\vq}+\tau[v_F^2\vq^2+(m+\lambda\Phi^s_\vq)^2]^\frac{1}{2}
\label{eq:eshift}
\eea
where the isotropy of the linear dispersive term (for small $|\vq|$) becomes obvious. Note, however that the parabolic and warping terms now depend manifestly on the magnetic field through the shift vector $\vk_s$ according to $E^s_{0\vq}=(\vq+\vk_s)^2/2m^*$ and $\Phi^s_\vq=\Phi_{\vq+\vk_s}= \bigl\{(q_x+k_{sx})[(q_x+k_{sx})^2-3(q_y+k_{sy})^2]\bigr\}$.\\
%
%%%%%%%%%%%%%%%%%%%%%%%%%%%%%%%%%%%%%%%%%%%%%%%%%%%%%%%%%%%%%
\begin{figure}
%\vspace{0.2cm}
\includegraphics[width=84mm]{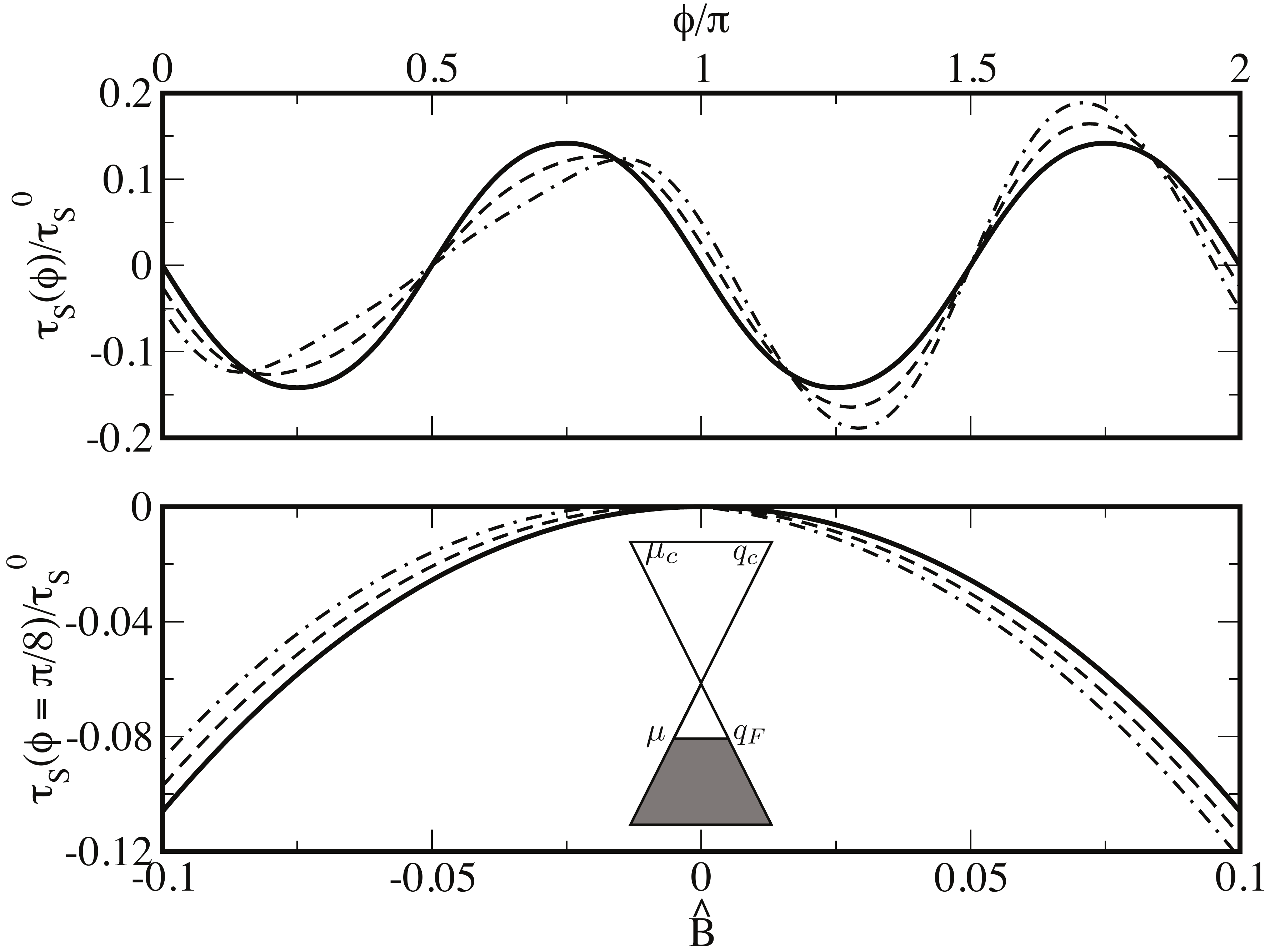}
\caption{Top: Torque oscillations as function of field angle $\phi$ for warping $\hl=1$. 
Full, dashed and dash-dotted lines correspond to $\ham=0, 0.025, 0.05$ respectively.
The symmetry of fourfold oscillations
$\tau_S(B,\phi+\pi)=\tau_S(B,\phi)$ which applies for $\ham = 0$ is lost for nonzero $\ham$.
Bottom: Low field behaviour of  torque at $\phi=\pi/8$ . Same line conventions as on top. For nonzero $\ham$ asymmetry 
$\tau_S(-\hB,\phi)\neq\tau_S(\hB,\phi)$ appears. Inset shows Dirac cones ($\ham =0$) with chemical potential position.}
\label{fig:Fig3}
\end{figure}
%%%%%%%%%%%%%%%%%%%%%%%%%%%%%%%%%%%%%%%%%%%%%%%%%%%%%%%%%%%%%
%

For the torque effect we need to calculate the paramagnetic susceptibility associated with the above surface state spectrum which is defined by
\bea
\chi_{\alpha\beta}^S=-\Bigl(\frac{\partial M^S_\alpha}{\partial B_\beta}\Bigr)
=-\Bigl(\frac{\partial^2 U_S}{\partial B_\alpha\partial B_\beta}\Bigr)
\label{eq:chidef}
\eea
where $U_S$ is the contribution of surface states to the internal energy per site. According to Eq.~(\ref{eq:eshift}) it has two contributions
\bea
U_S=U_0+U_D=\frac{1}{N}\sum_{\vq\tau}n_{\vq\tau}E^s_{0\vq}+
\frac{1}{N}\sum_{\vq\tau}\tau n_{\vq\tau}\epsilon^s_{\vq\tau}
\eea
with $n_{\vq\tau}=\theta_H(\mu-\epsilon^s_{\vq\tau})$ giving the occupation number of the surface state and $\mu$ denoting the chemical potential which lies in the bulk gap. It should be stressed that the integration runs over the occupied and rigidly shifted Dirac states up to a cuttoff wave number  $q_c=\mu_c/v_F$ where the surface states merge into the bulk. This rigid shift model is an approximation that does not take into account the details how the surface states merge into bulk at the cutoff and how this depends on the field induced shift of the origin. The parabolic term $U_0$ leads to a constant contribution $\chi^0_{\alpha\beta}(B)=-(\gamma_S^2/v_F^2m^*)n_S(\mu)\delta_{\alpha\beta}$ ($n_S$ = number of occupied surface states). It is proportional to the unit matrix and therefore does not contribute to the torque according to Eq.~(\ref{eq:torque1}). Therefore, in this equation we may replace  $\chi_{\alpha\beta}^S\rightarrow \chi_{\alpha\beta}^D$ since only the Dirac part $\chi_{\alpha\beta}^D$ may lead to a nonzero torque oscillation which is considered now.
The leading part of the Dirac cone contribution per site due to $U_D$ is given by
\bea
\chi^D_{\alpha\beta}=
-\frac{1}{N}\sum_\vq\Bigl( \frac{\partial^2\epsilon^s_\vq(\vB)}{\partial B_\alpha\partial B_\beta}\Bigr)
\sum_\tau\tau\theta_H(\mu-\epsilon^s_{\vq\tau})
\label{eq:suszD1}
\eea
From now on we assume particle hole symmetry, i.e. neglect $E_0$ in the Heaviside function $\theta_H$ of the above equation. 
Using $\sum_\tau\tau\theta_H(\mu-\tau\epsilon_\vq)=\theta_H(\mu-\epsilon_\vq)-\theta_H(\mu+\epsilon_\vq)$ one then obtains ($A_c=A/N$ is the surface unit cell area)
\bea
\label{eq:chi}
\chi^D_{\alpha\beta}
&=&\frac{A_c}{(2\pi)^2}\int_0^{2\pi}d\theta\int_{q_F}^{q_c}
dq q \frac{\partial^2\epsilon^s_\vq(\vB)}{\partial B_\alpha\partial B_\beta}
\label{eq:suszD2}
\eea
Note that this integral extends over the occupied part (shaded area in the inset of Fig.~\ref{fig:Fig3}) for $\mu < 0$. For $\mu >0$ due to particle-hole symmetry the contributions from $|\epsilon^s_{\vq\tau}|<\mu$ cancel and therefore $\chi^D_{\alpha\beta}(\mu)=\chi^D_{\alpha\beta}(-\mu)$ holds. For \BX~ the cutoff $q_c$ is given by $q_c/q_0\simeq 0.1$ with $q_0$ denoting the zone boundary wave vector. The field derivatives are only nonzero due to the existence of the warping term in Eq.~(\ref{eq:eshift}) and they are given by
\bea
\frac{\partial^2\epsilon^s_\vq(\vB)}{\partial B_\alpha\partial B_\beta}=\frac{\lambda^2}{\epsilon^s_\vq}
\Bigl[   
\bigl(\frac{\partial \Phi^s_\vq}{\partial B_\alpha}\bigr)\bigl( \frac{\partial \Phi^s_\vq}{\partial B_\beta}\bigr)
+\Phi^s_\vq \bigl(\frac{\partial^2 \Phi^s_\vq}{\partial B_\alpha\partial B_\beta}\bigr)
\Bigr]
\label{eq:deriv}
\eea
Evaluating the field derivatives of the warping function and  defining  $\Phi_\vq^s=q_c^3\hP^s_\vq$,  
$\bigl(\frac{\partial \Phi^s_\vq}{\partial B_\alpha}\bigr)=(\gamma_S/v_F)q^2_c\hP^s_{\hq\alpha}$ and
$\bigl(\frac{\partial^2 \Phi^s_\vq}{\partial B_\alpha\partial B_\beta}\bigr)=(\gamma_S/v_F)^2q_c\hP^s_{\hq\alpha\beta}$
(see Appendix \ref{sect:app}) one finally obtains the dimensionless susceptibility tensor of surface states as
%
%\begin{widetext}
\bea
\label{eq:susex}
\hat{\chi}^D_{\alpha\beta}=
\hl^2\int_0^{2\pi}d\theta\int_{\hq_F}^1d\hq
\frac{\hq[\hP_{\hq\alpha}^s \hP_{\hq\beta}^s+ \hP^s_{\hq} \hP^s_{\hq\alpha\beta}]}
{[\hq^2+(\ham+\hl\hP^s_{\vq})^2]^\frac{1}{2}}
\label{eq:suszD3}
\eea
%\end{widetext}
%
Here we introduced the dimensionless quantities  
$\hq=q/q_c$, $\ham=m/\mu_c$, $\hl=(\lambda/v_F)q_c^2=\lambda q_c^3/\mu_c$, $\hq_F=(\hmu^2-\ham^2)^\frac{1}{2}$ and $\hB=\gamma_SB/\mu_c$. Obviously 
$\hat{\chi}^D_{\alpha\beta}$ is only nonzero for finite warping parameter $\hl$. We define the amplitudes of
the two torque contributions  (Eq.~(\ref{eq:torque1}))  as $A_1(\phi)=\Delta\hat{\chi}^D=\hat{\chi}^D_{xx}-\hat{\chi}^D_{yy}$ and  $A_2(\phi)=2\hat{\chi}^D_{xy}$. \\

It is useful to consider first the low-field limit. For $B =0$ the integral in Eq.~(\ref{eq:susex}) reduces to diagonal form $\hat{\chi}^D_{\alpha\beta }=\hat{\chi}^D\delta_{\alpha\beta}$ with
\bea
\hat{\chi}^D=\chi^D/\chi_0^D=9\pi\hl^2\int_{\hq_F}^1\frac{\hq^5d\hq}{(\hq^2+\ham^2)^\frac{1}{2}}
\label{eq:suszD4}
\eea
where $\chi^D_0=\mu_c^{-1}\gamma_S^2(q_c/q_0)^2$. Therefore the zero-field susceptibility which vanishes identically for the isotropic Dirac spectrum becomes finite due to the effect of the warping. For $\ham =0$ one has $\hat{\chi}^D/\hat{\chi}_0^D=(9/5)\pi\hl^2(1-\hat{\mu}^5)$ which is almost flat for $\hmu\leq 0.4$ and the drops to zero at $\hmu =1$.
However, as in the case of the parabolic $\chi_{\alpha\beta}^0(B)$ contribution $\chi^D_{\alpha\beta}$(B=0) is also proportional to the unit matrix. As a result the zero field torque amplitudes $A_{1,2}(\phi)$ vanish meaning that the leading terms in $A_{1,2}(\phi)$ are of order $\sim B^2$. Since in additon  $\tau_S^0\sim B^2$ we get $\tau_S(\phi)\sim B^4$ for the low field behavior for the total torque. 
Explicitly, using Eqs.~(\ref{eq:torque1},\ref{eq:suszD3}) the magnetic torque is then given by
\bea
\hat{\tau}_S(\phi)=\frac{\tau_S(\phi)}{\tau_S^0}
&=&\Delta\hat{\chi}^D(\phi)\sin 2\phi - 2\hat{\chi}^D_{xy}(\phi)\cos 2\phi\no\\
\tau_S^0&=&\frac{1}{2}V_{SL}\mu_c^{-1}(\gamma_SB)^2(k_c/k_0)^2
\label{eq:osc}
\eea
Here the overall amplitude $\tau_S^0$ is determined by the volume $V_{SL}$ of the surface layer, the Zeeman energy and the ratio of the Dirac cone cross section at $\mu_c$ to the area of the surface BZ.

Furthermore the torque amplitudes have to fulfill symmetry requirements. In the massless case ($\ham =0$)  Eq.~(\ref{eq:susex}) and Eqs.~(\ref{eq:deriv1},\ref{eq:deriv2})  lead to $A_{1,2}(\phi+\pi)=A_{1,2}(\phi)$ and consequently to $\tau_S(B,\phi+\pi)=\tau_S(B,\phi)$. This is in accordance with the twofold symmetry of the underlying surface state dispersion 
(Fig.~\ref{fig:Fig1}c) around the shifted origin. For nonzero $\ham$ this symmetry is lost (Fig.~\ref{fig:Fig1}d) for the dispersion and therefore also for  $\tau_S(\phi)$ as shown later in Fig.~\ref{fig:Fig3}. However, by definition the relation $\tau_S(-B,\phi+\pi)=\tau_S(B,\phi)$ still holds. 

%
%%%%%%%%%%%%%%%%%%%%%%%%%%%%%%%%%%%%%%%%
\begin{center}
\begin{table}[b]
\caption{Typical Dirac cone parameters for \BX (Ref.~\onlinecite{zhou:09}). 
For the torque amplitudes the length
and energy scales $q_c^{-1}$ and $\mu_c$ are used. The intrinsic length and
energy scales $b$ and  $E^*$  associated with the warping $\lambda$ are only 
used for Fig.~\ref{fig:Fig1} .}
\vspace{0.2cm}
\begin{center}
\begin{tabular}{l @{\hspace{4mm}} l @{\hspace{4mm}} l }
\hline\hline
Fermi velocity&  $v_F$ & 2.55 eV \AA\\
cutoff wave number & $q_c$ & 0.10 \AA$^{-1}$ \\
cutoff energy & $\mu_c=v_Fq_c$ & 0.255 eV  \\
intrinisic energy scale & $E^*= (v_F/b)$   & 0.25 eV\\
intrinsic length scale & $b=(\lambda/v_F)^\frac{1}{2}$ & $10.2$ \AA \\
warping parameter &$\lambda=v_F^3/E^{*2}$ &  $2.65\cdot 10^2$ eV \AA$^3$\\
(dimensionless) & $\hl=\lambda q_c^3/\mu_c$ & 1.04\\
surface state g-factor& $g_S$ & 52\\ 
field energy scale & $\gamma_S=\frac{1}{2}g_S\mu_B$ & $1.44$ meV/T\\
\hline\hline
\end{tabular}
\end{center}
\label{table:tableI}
\end{table}
\end{center}
%%%%%%%%%%%%%%%%%%%%%%%%%%%%%%%%%%%%%%%%%%
%
For discussion of numerical results obtained from Eqs.~(\ref{eq:susex},\ref{eq:osc}) we give the typical magnitude of Dirac cone parameters for \BX~ in Table \ref{table:tableI}. Using values of $\mu_c$ and $\gamma_S$ given there the reasonably accessible field range is $ 0 \leq \hB \leq 0.1$ where the upper value corresponds to B = 17.8 T.

\section{Numerical results and discussion}
\label{sect:numerical}

The surface state dispersion given in Eq.~(\ref{eq:dispk}) is the fundamental property that determines the torque oscillations. It is shown in Fig.~\ref{fig:Fig1}a-d (clockwise direction) by contour plots for different values of mass $\ham$ and field $\hB$ . There we used the intrinsic length scale $b$ and energy scale $E^*$ related to the warping parameter \cite{zhou:09}. For ($\tilde{m},\tilde{B}$)= (0,0) we obtain the typical snowflake shape  of the warped ungapped Dirac cones (a) that has been observed in photoemission \cite{chen:09}. It exhibits both twofold and threefold rotational symmetry. The principal effect of the mass term is to remove these symmetries as seen in (b) where for clarity a large mass was chosen. The effect of the paramagnetic term due to an in-plane field and for zero mass is shown in (c): For finite B $(\phi=3\pi/4)$ the center of the Dirac cone is now shifted from the origin of the BZ to the point $(bk_x, bk_y)=\tB(\sin\phi,-\cos\phi)=\frac{\sqrt{2}}{2}\tB(1,1)$ but the rotational symmetry around the shifted center is preserved at low energy. The direction of the shift depends on the direction of the field (chosen along a diagonal here). For finite mass and field both effects (symmetry reduction and shift) are combined (d).

The dispersion enters directly in the denominator of the susceptibility tensor in Eq.~(\ref{eq:susex}) which determines the torque amplitudes $A_{1,2}(\phi)$. In Fig.~\ref{fig:Fig2} they are first plotted as function of the warping parameter for a field angle $\phi$ where both $A_{1,2}$  give non-zero contribution to $\tau_S(\phi)$. As can be seen directly from Eq.~(\ref{eq:susex}) the amplitudes are invariant under $\hl\rightarrow -\hl$ in the massless case (top) but not for finite $\ham$. This applies also for the total torque $\hat{\tau}_S$ (bottom). The $A_{1,2}$ amplitudes are nonmonotonic in $\hl$, however the total $|\hat{\tau}_S(\phi)|$ still increases monotonically with $|\hl|$. For small warping and general $\phi$ $\hat{\tau}_S(\phi)\sim\hl^2$ coming from the field derivatives of the warping function and for larger $|\hl|$ $\hat{\tau}_S(\phi)$ increases rapidly due to the effect of the warped dispersion in the denominator of Eq.~(\ref{eq:suszD3}). The value relevant for \BX~ is close to the intermediate value $\hl=1$ which will be used in the following figures.

%
%%%%%%%%%%%%%%%%%%%%%%%%%%%%%%%%%%%%%%%%%%%%%%%%%%%%%%%%%%%%%
\begin{figure}
%\vspace{0.2cm}
\includegraphics[width=84mm]{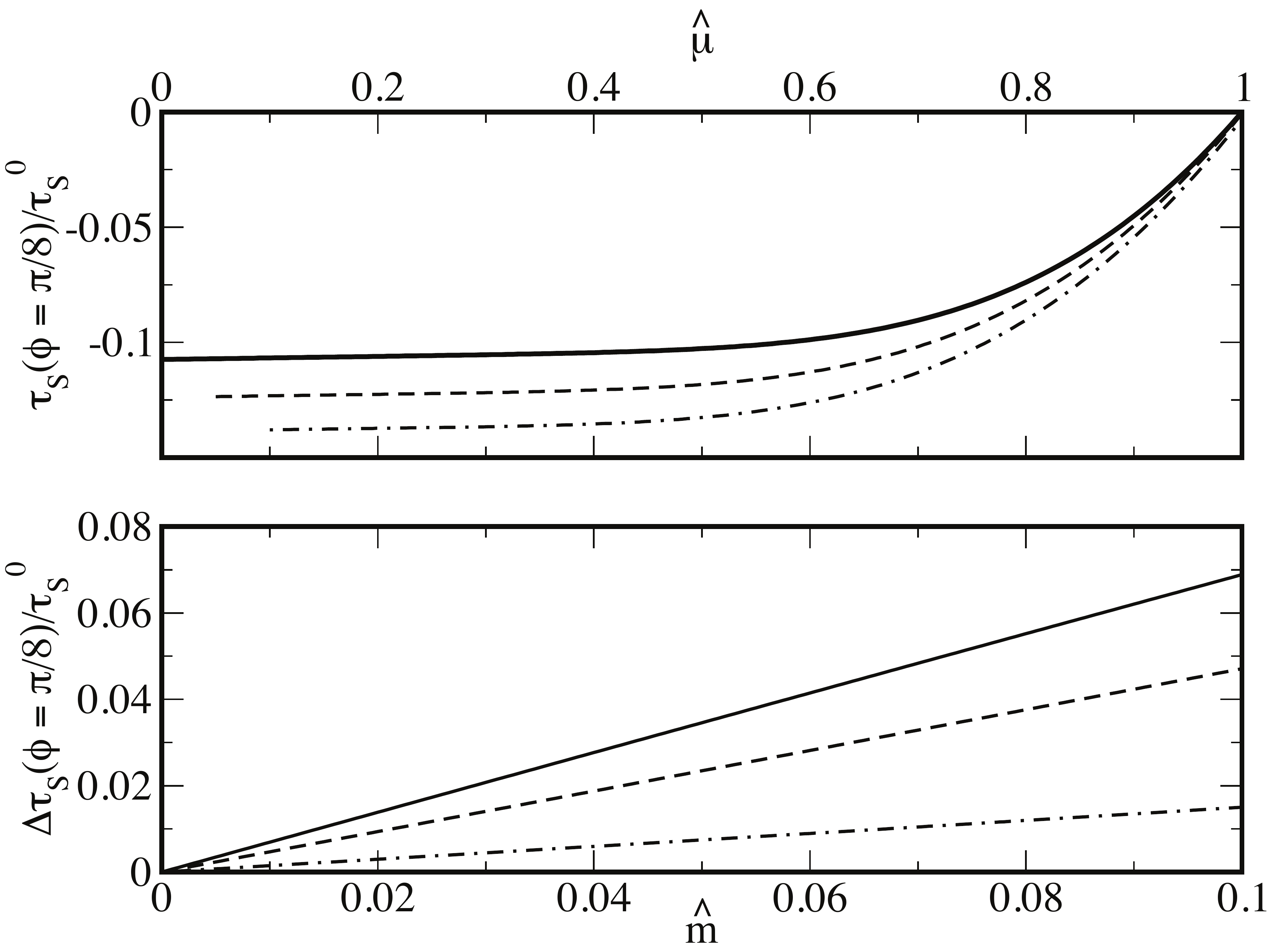}
\caption{Top: Variation of $\phi=\pi/8$ torque with chemical potential ($\hat{\mu}=\mu/\mu_c$) for mass gap $\ham =0, 0.05, 0.1$ (full, dashed, dash-dotted lines respectively). Due to particle-hole symmetry $\hat{\tau}_S(\hmu)=\hat{\tau}_S(-\hmu)$.
The minimum chemical potential is $|\hmu| = \ham$. For $|\hat{\mu}|=1$ the Fermi level lies on the bottom of conduction or top of valence bands and the surface contribution $\hat{\tau}_S$ vanishes.
Bottom: Torque asymmetry $\Delta\hat{\tau}_S(B,\phi)=\hat{\tau}_S(B,\phi+\pi)-\hat{\tau}_S(B,\phi)$ as function of the mass gap for
$|\hat{\mu}|=0.2, 0.7, 0.9$  (full, dashed, dash-dotted lines respectively)}
\label{fig:Fig4}
\end{figure}
%%%%%%%%%%%%%%%%%%%%%%%%%%%%%%%%%%%%%%%%%%%%%%%%%%%%%%%%%%%%%
%

The actual torque oscillations with field angle are shown in  Fig.~\ref{fig:Fig3} (top) for $\hB = 0.1$. They exhibit fourfold symmetry in the massless case (thick line) which is broken for finite mass. For the latter therefore a nonzero asymmetry $\Delta\tau_S(B,\phi)=\tau_S(B,\phi+\pi)-\tau_S(B,\phi)$ appears. As can be seen in  Fig.~\ref{fig:Fig3} (bottom) $\hat{\tau}_S$ varies quadratically with $\hB$. This figure directly shows the asymmetry of the torque under $\phi\rightarrow\phi+\pi$ or $\hB\rightarrow -\hB$ for $\ham\neq 0$ (broken lines).

In Fig.~\ref{fig:Fig4} we show the extremal torque amplitude as function of chemical potential. The torque oscillations become maximal when $\hat{\mu}\rightarrow 0$ for half occupancy of the surface states. We remind that for $\hmu > 0$ (chemical potential in the upper half cone in Fig.\ref{fig:Fig3}b) there is partial cancellation of contributions due to particle-hole symmetry and therefore $\tau_S(\hat{\mu})=\tau_S(-\hat{\mu})$. The states close to the Dirac point show little warping, therefore when $\hmu$ crosses that region the torque amplitude changes slowly. This is due to the warping function appearing in the numerator of Eq.~(\ref{eq:suszD3}). The strong variation of the torque for larger chemical potential in Fig.~\ref{fig:Fig4}(top)  would be the signature of its surface state origin in an experiment. This may possibly be achieved by an additional gate voltage tuning on the disk as demonstrated in Ref.~\onlinecite{chen:10b}.\\

\section{Conclusion and outlook}
\label{sect:conclusion}

We have calculated the magnetic torque expected from helical surface states of a topological insulator when the magnetic field
is rotated in the surface plane. For finite warping of the Dirac cones the paramagnetic moment induced in the surface states is non-collinear with the field and leads to torque oscillations. The  total amplitude increases with fourth power in the field. Furthermore the presence of a mass gap in the dispersion leads to a  symmetry reduction of the fourfold oscillations  which should be detectable experimentally. The surface torque also depends considerably on the chemical potential position in the bulk band gap sufficiently away from the Dirac point, in contrast to the torque originating from the bulk states. Furthermore the ratio of the two contributions may be systematically changed by varying the layer numbers in thin film samples. These two effects should facilitate the separation of the surface torque from the constant bulk background.  In this work we considered only constant and uniform fields. It seems possible to extend the theory to the case of  dynamical and nonuniform torque using appropriate spin response functions \cite{garate:10}.

\appendix
\section{}
\label{sect:app}

Here we give the explicit form of dimensionless warping function $\hP^s_{\hq}$ and its field derivatives $\hP^s_{\hq\alpha}, \hP^s_{\hq\alpha\beta}$  that appear in the numerator 
of the susceptibility tensor $\hat{\chi}^S_{\alpha\beta}$ in Eq.~(\ref{eq:suszD3}). Using polar representation of momentum $\vq=q(\cos\theta,\sin\theta)$  and field $\vB=B(\cos\phi,\sin\phi)$ coordinates we obtain for the first order derivatives 
\bea
\label{eq:deriv1}
\hP^s_{\hq}&=&(\hq\cos\theta+\hB\sin\phi)\times\\
&&\bigl[(\hq\cos\theta+\hB\sin\phi)^2-3(\hq\sin\theta-\hB\cos\phi)^2\bigr]\no\\
\hP^s_{\hq x}&=&3\bigl[\hq^2\sin 2\theta - \hB^2\sin 2\phi -2\hq\hB\cos(\theta+\phi)\bigr]\no\\
\hP^s_{\hq y}&=&3\bigl[\hq^2\cos 2\theta - \hB^2\cos 2\phi -2\hq\hB\sin(\theta+\phi)\bigr]\no
\eea
and likewise for the second order field derivatives
\bea
-\hP^s_{\hq xx}=\hP^s_{\hq yy}&=&6(\hq\cos\theta+\hB\sin\phi)\no\\
\hP^s_{\hq xy}=\hP^s_{\hq yx}&=&6(\hq\sin\theta-\hB\cos\phi)
\label{eq:deriv2}
\eea
In the case B=0 relevant for the zero-field susceptibility in Eq.~(\ref{eq:suszD4}) this reduces to
\bea
(\hP^s_{\hq})_0=\hP_{\hq}&=&\hq^3\cos 3\theta\no\\
\hP^s_{\hq x}&=&3\hq^2\sin 2\theta\no\\
\hP^s_{\hq y}&=&3\hq^2\cos 2\theta\\
-\hP^s_{\hq xx}=\hP^s_{\hq yy}&=&6\hq\cos\theta\no\\
\hP^s_{\hq xy}=\hP^s_{\hq yx}&=&6\hq\sin\theta\no
\label{eq:deriv3}
\eea

\end{document}